\documentclass[aps,prl,twocolumn,tightenlines,superscriptaddress,showpacs,preprintnumbers,amsmath,amssymb]{revtex4}
\usepackage{graphicx} % Include figure files
\usepackage{dcolumn}  % Align table columns on decimal point
\usepackage{bm}% bold math
\usepackage{pstricks}% required
\usepackage{pst-node}% required

\graphicspath{{ps}}

\usepackage{subfigure}
\usepackage{mathrsfs}
\usepackage{amssymb}
\usepackage{amsmath}
\usepackage{tabularx}
\usepackage{rotating}
\usepackage{multirow}

\newcommand{\bz}{{B^0}}
\newcommand{\bzb}{{\overline{B}{}^0}}

\newcommand{\dE}{{\Delta E}}
\newcommand{\mb}{{M_{\rm bc}}}
\newcommand{\mbc}{{M_{\rm bc}}}

\newcommand{\ftag}{f_{\rm tag}}

\newcommand{\ttag}{t_{\rm tag}}

\newcommand{\dm}{\Delta m_d}
\newcommand{\dmd}{\dm}
\newcommand{\taubz}{{\tau_\bz}}

\newcommand*{\dwl}{\ensuremath{{\Delta w_l}}}

\newcommand{\pip}{{\pi^+}}
\newcommand{\pim}{{\pi^-}}
\newcommand{\piz}{{\pi^0}}

\newcommand{\lsig}{{\cal L}_{\rm sig}}
\newcommand{\lbkg}{{\cal L}_{\rm bkg}}
\newcommand{\rsigbkg}{{\cal R}}

\newcommand{\nbb}{449}

\newcommand{\lint}{414}

\newcommand{\kakkoOverlineBeta}{\raise1ex\hbox{\scriptsize ${}^($}
  \overline{\beta} \raise1ex\hbox{\scriptsize ${}^)$} \! \!}
\newcommand{\kakkoOverlinef}{\raise1ex\hbox{\scriptsize ${}^($}
  \overline{f} \raise1ex\hbox{\scriptsize ${}^{)}$} \! \!}
\newcommand{\kakkoOverlineF}{\raise1ex\hbox{\scriptsize ${}^($}
  \overline{F} \raise1ex\hbox{\scriptsize ${}^{)}$} \! \!}
\newcommand{\kakkoOverlineA}{\raise1ex\hbox{\scriptsize ${}^($}
  \overline{A} \raise1ex\hbox{\scriptsize ${}^{)}$} \! \!}
\newcommand{\kakkoOverlineGamma}{{\raise0.5ex\hbox{\scriptsize ${}^($}
  \overline{\gamma} \raise0.5ex\hbox{\scriptsize ${}^)$} \! \!}}

\newcommand{\pipipi}{\pi^+\pi^-\pi^0}

\begin{document}

%\vspace*{-3\baselineskip}
%\resizebox{!}{3cm}{\includegraphics{belle.eps}}

\preprint{\vbox{ \hbox{   }
                        \hbox{Belle Preprint 2007-4}
                        \hbox{KEK Preprint 2006-65}
}}

\title{ \quad\\[0.5cm]  
\boldmath Measurement of $CP$ Asymmetry in a Time-Dependent Dalitz
Analysis
of $B^0 \to (\rho \pi)^0$ and a Constraint on the Quark Mixing Matrix
Angle $\phi_2$}

%%%% >>>>> insert the authorlist here. BEFORE the abstract !!!!! <<<<<
%%%% >>>>> from the authorship confirmation web page
%%% Name the file author.tex and use \input{author} to insert into your latex file.
%\author{Author}%\affiliation{affiliation}
%%% INPUT_EXPANDED(authors.tex) BEGIN %%%
\affiliation{Budker Institute of Nuclear Physics, Novosibirsk}
\affiliation{Chiba University, Chiba}
\affiliation{University of Cincinnati, Cincinnati, Ohio 45221}
\affiliation{Department of Physics, Fu Jen Catholic University, Taipei}
\affiliation{The Graduate University for Advanced Studies, Hayama, Japan}
\affiliation{Gyeongsang National University, Chinju}
\affiliation{Hanyang University, Seoul}
\affiliation{University of Hawaii, Honolulu, Hawaii 96822}
\affiliation{High Energy Accelerator Research Organization (KEK), Tsukuba}
\affiliation{University of Illinois at Urbana-Champaign, Urbana, Illinois 61801}
\affiliation{Institute of High Energy Physics, Chinese Academy of Sciences, Beijing}
\affiliation{Institute of High Energy Physics, Vienna}
\affiliation{Institute of High Energy Physics, Protvino}
\affiliation{Institute for Theoretical and Experimental Physics, Moscow}
\affiliation{J. Stefan Institute, Ljubljana}
\affiliation{Kanagawa University, Yokohama}
\affiliation{Korea University, Seoul}
\affiliation{Kyungpook National University, Taegu}
\affiliation{Swiss Federal Institute of Technology of Lausanne, EPFL, Lausanne}
\affiliation{University of Ljubljana, Ljubljana}
\affiliation{University of Maribor, Maribor}
\affiliation{University of Melbourne, Victoria}
\affiliation{Nagoya University, Nagoya}
\affiliation{Nara Women's University, Nara}
\affiliation{National Central University, Chung-li}
\affiliation{National United University, Miao Li}
\affiliation{Department of Physics, National Taiwan University, Taipei}
\affiliation{H. Niewodniczanski Institute of Nuclear Physics, Krakow}
\affiliation{Nippon Dental University, Niigata}
\affiliation{Niigata University, Niigata}
\affiliation{University of Nova Gorica, Nova Gorica}
\affiliation{Osaka City University, Osaka}
\affiliation{Osaka University, Osaka}
\affiliation{Panjab University, Chandigarh}
\affiliation{Peking University, Beijing}
\affiliation{Princeton University, Princeton, New Jersey 08544}
\affiliation{RIKEN BNL Research Center, Upton, New York 11973}
\affiliation{University of Science and Technology of China, Hefei}
\affiliation{Seoul National University, Seoul}
\affiliation{Shinshu University, Nagano}
\affiliation{Sungkyunkwan University, Suwon}
\affiliation{University of Sydney, Sydney NSW}
\affiliation{Tata Institute of Fundamental Research, Bombay}
\affiliation{Toho University, Funabashi}
\affiliation{Tohoku Gakuin University, Tagajo}
\affiliation{Tohoku University, Sendai}
\affiliation{Department of Physics, University of Tokyo, Tokyo}
\affiliation{Tokyo Institute of Technology, Tokyo}
\affiliation{Tokyo Metropolitan University, Tokyo}
\affiliation{Tokyo University of Agriculture and Technology, Tokyo}
\affiliation{Virginia Polytechnic Institute and State University, Blacksburg, Virginia 24061}
\affiliation{Yonsei University, Seoul}
 \author{A.~Kusaka}\affiliation{Department of Physics, University of Tokyo, Tokyo} % Tokyo
 \author{C.~C.~Wang}\affiliation{Department of Physics, National Taiwan University, Taipei} % Taiwan
 \author{H.~Ishino}\affiliation{Tokyo Institute of Technology, Tokyo} % TIT
 \author{K.~Abe}\affiliation{High Energy Accelerator Research Organization (KEK), Tsukuba} % KEK
 \author{K.~Abe}\affiliation{Tohoku Gakuin University, Tagajo} % TohokuGakuin
 \author{I.~Adachi}\affiliation{High Energy Accelerator Research Organization (KEK), Tsukuba} % KEK
 \author{H.~Aihara}\affiliation{Department of Physics, University of Tokyo, Tokyo} % Tokyo
 \author{D.~Anipko}\affiliation{Budker Institute of Nuclear Physics, Novosibirsk} % BINP
 \author{V.~Aulchenko}\affiliation{Budker Institute of Nuclear Physics, Novosibirsk} % BINP
 \author{T.~Aushev}\affiliation{Swiss Federal Institute of Technology of Lausanne, EPFL, Lausanne}\affiliation{Institute for Theoretical and Experimental Physics, Moscow} % ITEP
% ...
% Ljubljana
\collaboration{The Belle Collaboration} 

%%% INPUT_EXPANDED(authors.tex) END %%%%%
% \collaboration{The Belle Collaboration}
\noaffiliation
%% end author list

\begin{abstract}
We present a measurement of $CP$ asymmetry using a time-dependent
Dalitz plot analysis of
$B^0 \to \pipipi$ decays based on a $\lint \, {\rm fb}^{-1}$ data sample 
containing $\nbb\times 10^6 B\overline{B}$ pairs.
The data was collected 
on the $\Upsilon(4S)$ resonance
with the Belle detector at the KEKB asymmetric energy $e^+ e^-$
collider.
Combining our analysis with information on charged $B$ decay modes,
we perform a full Dalitz and isospin analysis
 and obtain a constraint on the CKM angle $\phi_2$,
$68^\circ < \phi_2 < 95^\circ$ as the 68.3\% confidence interval
for the $\phi_2$ solution consistent with the standard model (SM).
A large SM-disfavored region also remains.
\end{abstract}


\maketitle

%%%% >>>> keep the final version single-spaced
%\tighten

{\renewcommand{\thefootnote}{\fnsymbol{footnote}}}
\setcounter{footnote}{0}

% ============================================================
% Introduction
% ============================================================
In the standard model (SM),
 $CP$ violation arises from an irreducible phase in the 
 Cabibbo-Kobayashi-Maskawa (CKM)
 matrix~\cite{Cabibbo:1963yz,Kobayashi:1973fv}.
% The SM predicts that measurement of a $CP$ asymmetry
%  in time-dependent decay rates of $\bz$ and $\bzb$
%  gives access to the $CP$ violating phase
%  in the CKM matrix~\cite{Carter:1980hr,Carter:1980tk,Bigi:1981qs}.
% The angle $\phi_2$ of the CKM unitarity triangle
%  can be measured via
% the tree diagram contribution in
% $b\rightarrow u \overline{u} d$ decay processes.
% In the decay processes of this category, however,
%  contributions from so-called penguin diagrams
%  could contaminate the measurement of $\phi_2$.
%
Snyder and Quinn pointed out that
 a time-dependent Dalitz plot analysis (TDPA)
 of the decay $B^0 \to \rho\pi \to \pipipi$~\cite{ChargeConjugate}
 offers a unique way to determine the angle $\phi_2$~\cite{Phi2Alpha}
 in the CKM
 unitarity triangle without discrete ambiguities,
% (for $\phi_2$ in the range between 0 and $\pi$),
 which cannot be obtained from analyses of
 other modes sensitive to $\phi_2$ such as $B \to \pi\pi$ or
 $\rho\rho$~\cite{Snyder:1993mx}.
% taking account of possible
% The time-dependent Dalitz plot analysis
 The TDPA
 uses isospin and takes into
 account a possible
 contamination from $b\rightarrow d$ penguin transitions.
In addition, using measurements of 
% the related charged decay modes
 $B^+\rightarrow \rho^+\pi^0$ and $\rho^0\pi^+$
 provides further improvement of the $\phi_2$
 determination~\cite{Lipkin:1991st,Gronau:1991dq}.

% ============================================================
% Detector and Data Set
% ============================================================
In this Letter, we present the result of
% a time-dependent Dalitz plot analysis
 a TDPA
 in $\bz \to \pipipi$ decays and a constraint on $\phi_2$.
We use a $\lint \, {\rm fb}^{-1}$ data sample 
 that contains $\nbb\times 10^6 B\overline{B}$ pairs collected 
 on the $\Upsilon(4S)$ resonance.
The data were taken at the KEKB collider~\cite{KEKB}
 using the Belle detector~\cite{Belle}.

In the decay chain $\Upsilon(4S) \to \bz \bzb \to (\pipipi) \ftag$,
 where
%  one of the $B$ mesons decays at time $\tCP$
%  to a final state $\fCP = \pipipi$
%  and the other decays at time $\ttag$ to a final state $\ftag$
%  that distinguishes $\bz$ and $\bzb$,
 $\ftag$ is a final state 
 that distinguishes $\bz$ and $\bzb$,
 the time- and Dalitz plot-dependent differential decay rate is
\begin{widetext}
\begin{equation}
 \frac{d \Gamma}{d \Delta t \, ds_+ ds_-} \sim
   e^{- |\Delta t| / \taubz}  \Bigl\{
   \left(|A_{3\pi}|^2 + |\overline{A}_{3\pi}|^2\right)
   - q_\mathrm{tag} \cdot
    \left(|A_{3\pi}|^2 - |\overline{A}_{3\pi}|^2\right)
    \cos (\dmd \Delta t)
    + q_\mathrm{tag} \cdot
     2 \mathrm{Im}\left[
   		     \frac{q}{p}
   		     A_{3\pi}^*
   		     \overline{A}_{3\pi}
   		    \right]
    \sin (\dmd \Delta t)
    \Bigr\}
    \;.
 \label{equ:amplitude_dt_width}
\end{equation}
\end{widetext}
Here, $\kakkoOverlineA {}_{3\pi}$
 is the Lorentz-invariant
 amplitude of the $\bz(\bzb) \to \pipipi$ decay,
 $q_\mathrm{tag}$ is the $b$-flavor charge
 ($q_\mathrm{tag} = +1 \: (-1)$
 when $\ftag$ is a $\bz$ ($\bzb$) flavor eigenstate),
 % and $\Delta t \equiv \tCP - \ttag$;
 and $\Delta t$ is the decay time difference of the two $B$ mesons
 ($t_{3\pi} - \ttag$).
 The parameters
 $p$ and $q$ define the mass eigenstates of neutral $B$
 mesons as $p \bz \pm q \bzb$,
 with an average lifetime $\taubz$ and mass difference $\dm$.
The Dalitz plot variables $s_+$, $s_-$, and $s_0$
 are defined as
\begin{equation}
  s_+ \equiv (p_+ + p_0)^2 , \;
  s_- \equiv (p_- + p_0)^2 , \;
  s_0 \equiv (p_+ + p_-)^2 ,
\end{equation}
 where $p_+$, $p_-$, and $p_0$ are the four-momenta of the
 $\pi^+$, $\pi^-$, and $\pi^0$, respectively,
 in the decay of $\bz \to \pipipi$.

The amplitudes $\kakkoOverlineA {}_{3\pi}$ have
the following Dalitz plot dependences
\begin{eqnarray}
 A_{3\pi}(s_+, s_-) & = &
  \sum_{\kappa = (+, -, 0)} 
  f_\kappa(s_+, s_-) A^\kappa \;,
  % f_+(s_+, s_-) A^+ + f_-(s_+, s_-) A^- + f_0(s_+, s_-) A^0 \;,
  \label{equ:amplitude_q2b_b}
 \\
 \frac{q}{p}
  \overline{A}_{3\pi}(s_+, s_-)
  & = &
  \sum_{\kappa = (+, -, 0)} 
  \overline{f}{}_\kappa(s_+, s_-) \overline{A}{}^\kappa \;,
  % \overline{f}_+(s_+, s_-) \overline{A}{}^+
  % + \overline{f}_-(s_+, s_-) \overline{A}{}^-
  % + \overline{f}_0(s_+, s_-) \overline{A}{}^0 \;,
  \label{equ:amplitude_q2b_bbar}
\end{eqnarray}
 where $A^\kappa (\overline{A}{}^\kappa)$
 are complex amplitudes corresponding to
 $\bz(\bzb) \to \rho^+ \pi^-, \rho^- \pi^+, \rho^0 \pi^0$
 for $\kappa = +, -, 0$.
% $\bz(\bzb) \to \rho^\kappa \pi^{\overline{\kappa}}$
% with $(\overline{+}, \overline{-}, \overline{0}) \equiv (-, +, 0)$;
% $\rho^\kappa \pi^{\overline{\kappa}} = (\rho^+ \pi^-, \rho^- \pi^+,
% \rho^0 \pi^0)$ for $\kappa = (+, -, 0)$.
Here we neglect possible contributions
 to the $\bz \to \pipipi$ decay
 other than that of $\bz \to (\rho\pi)^0 \to \pipipi$
 and take account of them as systematic uncertainties.
%~\cite{KappaBarDefinition}.
The functions $\kakkoOverlinef {}_\kappa$
 incorporate the kinematic and dynamical properties of
%  the $\bz$
%  decaying into a vector $\rho^\kappa$ and
%  a pseudoscalar $\pi^{\overline{\kappa}}$.
%  They can be factorized into two parts as
%  $\bz \to \rho^\kappa \pi^{\overline{\kappa}}$
  $\bz \to (\rho\pi)^0$
  decays and can be written as
\begin{equation}
 \kakkoOverlinef {}_\kappa (s_+, s_-)
  = T_{J=1}^\kappa \, \kakkoOverlineF {}_{\pi}^{\: \kappa}(s_\kappa)
  \quad (\kappa = +, -, 0) \;,
\end{equation}
 where $T_{J=1}^\kappa$ and
 $\kakkoOverlineF {}_{\pi}^{\: \kappa}(s_\kappa)$
 correspond to the helicity distribution
 and the lineshape of $\rho^\kappa$, respectively.
The lineshape is parameterized with Breit-Wigner
 functions
%
% FOLLOWING_CAN_BE_OMITTED
%
 corresponding to the
% $\rho(770)$, $\rho(1450)$, and $\rho(1700)$ resonances:
 $\rho(770)$ and its radial excitations:
\begin{equation}
 \kakkoOverlineF {}_{\pi}^{\: \kappa} (s)
  =
  \mathrm{BW}_{\rho(770)}
  + \kakkoOverlineBeta {}_\kappa \, \mathrm{BW}_{\rho(1450)}
  + \kakkoOverlineGamma {}_\kappa \, \mathrm{BW}_{\rho(1700)} \;,
  \label{equ:fpi_lineshape_beta_gamma}
\end{equation}
 where the amplitudes $\kakkoOverlineBeta {}_\kappa$ and
 $\kakkoOverlineGamma {}_\kappa$
 (denoting the relative sizes of two resonances)
 are complex numbers.
We use the Gounaris-Sakurai (GS) model~\cite{Gounaris:1968mw}
 for the Breit-Wigner shape
 and the world average~\cite{Eidelman:2004wy} for the mass and width
 of each resonance.
Though $\kakkoOverlineBeta {}_\kappa$ and
 $\kakkoOverlineGamma {}_\kappa$ can be different
 for each of six decay modes of
% $\bz(\bzb) \to \rho^\kappa\pi^{\overline{\kappa}}$ in general,
 $\bz(\bzb) \to (\rho\pi)^0$ in general,
 we assume no such variation, i.e.,
$\kakkoOverlineBeta {}_\kappa = \beta$ and
$\kakkoOverlineGamma {}_\kappa = \gamma$,
% This assumption leads to a relation
% $\kakkoOverlineF {}_{\pi}^{\: \kappa} (s) = F_{\pi} (s)$
% and thus $\overline{f}_\kappa (s_+, s_-) = f_\kappa (s_+, s_-)$.
%
%  \begin{equation}
%   \kakkoOverlineF {}_{\pi}^{\: \kappa} (s)
%    = F_{\pi} (s)
%    \equiv 
%    BW_{\rho(770)}
%    + \beta \, BW_{\rho(1450)}
%    + \gamma \, BW_{\rho(1700)},
%    \label{equ:lineshape_no_variation_assumption}
%  \end{equation}
  in our nominal fit, and address
  possible deviations from this assumption in the systematic error.
% Equation (\ref{equ:lineshape_no_variation_assumption}) leads to
This assumption leads to
 the relation $\overline{f}_\kappa (s_+, s_-) = f_\kappa (s_+, s_-)$.
%
% \begin{equation}
%  \overline{f}_\kappa (s_+, s_-) = f_\kappa (s_+, s_-) 
%   \quad (\kappa = +, -, 0) \;.
%   \label{equ:f_kappa_equality_with_assumption}
% \end{equation}

With this relation and Eqs.~(\ref{equ:amplitude_q2b_b}) and
(\ref{equ:amplitude_q2b_bbar}),
the coefficients of Eq.~(\ref{equ:amplitude_dt_width}) are
\begin{widetext}
\begin{equation}
 |A_{3\pi}|^2 \pm |\overline{A}_{3\pi}|^2
  = \sum_{\kappa \in \{+,-,0\}} |f_\kappa|^2 U^\pm_\kappa
  + 2 \hspace{-4mm} \sum_{\kappa<\sigma \in \{+,-,0\}} \hspace{-3mm}
  \left(
   \mathrm{Re}[f_\kappa f^*_\sigma] U^{\pm,\mathrm{Re}}_{\kappa\sigma}
   - \mathrm{Im}[f_\kappa f^*_\sigma] U^{\pm,\mathrm{Im}}_{\kappa\sigma}
  \right)
  \;,
\end{equation}
\begin{equation}
 \mathrm{Im}\left(\frac{q}{p} A_{3\pi}^* \overline{A}_{3\pi} \right)
  = \sum_{\kappa \in \{+,-,0\}} |f_\kappa|^2 I_\kappa
  + \hspace{-4mm} \sum_{\kappa<\sigma \in \{+,-,0\}} \hspace{-3mm} 
  \left(
   \mathrm{Re}[f_\kappa f^*_\sigma] I^\mathrm{Im}_{\kappa\sigma}
   + \mathrm{Im}[f_\kappa f^*_\sigma] I^\mathrm{Re}_{\kappa\sigma}
  \right)
  \;,
\end{equation}
\end{widetext}
with
\begin{eqnarray}
 U^\pm_{\kappa}
  & = & |A^\kappa|^2 \pm |\overline{A}{}^\kappa|^2 \;,
  \label{equ:fit_params_first}
  \\
 I_\kappa & = & \mathrm{Im}
  \left[
   \overline{A}{}^{\kappa} A^{\kappa *}
  \right] \;,
  \label{equ:fit_params_2}
  \\
 U^{\pm,\mathrm{Re}(\mathrm{Im})}_{\kappa\sigma}
  & = & \mathrm{Re}(\mathrm{Im})
  \left[ A^\kappa A^{\sigma *} \pm \overline{A}{}^\kappa 
    \overline{A}{}^{\sigma *}
  \right] \;,
  \label{equ:fit_params_3}
  \\
 I^\mathrm{Re(Im)}_{\kappa \sigma} & = & \mathrm{Re(Im)}
  \left[
   \overline{A}{}^\kappa A^{\sigma *}
   \! - \! (+) \, \overline{A}{}^\sigma A^{\kappa *}
  \right] \;.
  \label{equ:fit_params_last}
%  \\
%  I^\mathrm{Im}_{\kappa \sigma} & = & \mathrm{Im}
%   \left[
%    \overline{A}{}^\kappa A^{\sigma *}
%    + \overline{A}{}^\sigma A^{\kappa *}
%   \right] \;.
%   \label{equ:fit_params_last}
\end{eqnarray}
The 27 coefficients
 (\ref{equ:fit_params_first})--(\ref{equ:fit_params_last})
 are the parameters determined by the fit~\cite{Quinn:2000by}.
The parameters (\ref{equ:fit_params_first})--(\ref{equ:fit_params_2})
and (\ref{equ:fit_params_3})--(\ref{equ:fit_params_last})
are called
non-interfering and interfering parameters, respectively.
This parameterization allows us to describe the differential
decay width as a linear combination of independent functions,
whose coefficients are fit parameters
in a well behaved fit.
We fix the overall normalization by requiring $U^+_+ = 1$.
Thus, 26 of the 27 coefficients are free parameters in the fit.

% (Candidate 1)
% [They uniquely determine all the relative
% sizes and phases of the amplitudes $A^\kappa$ and $\overline{A}{}^\kappa$,
% which are related to $\phi_2$ through an isospin
%  relation~\cite{Lipkin:1991st,Gronau:1991dq} by
% \begin{equation}
%  e^{+2i\phi_2} = 
%   \frac{\overline{A}{}^+ + \overline{A}{}^- + 2 \overline{A}{}^0}
%   {A^+ + A^- + 2 A^0} \;.
% \end{equation}
% % They in principle allow us to
% Consequently, we can constrain $\phi_2$ without
% discrete ambiguity.
% Here, the uniqueness of this analysis comes from the measurement of the
% interfering parameters, which corresponds to the measurements of
% $CP$-violating asymmetries in mixed final states and plays an essential
% role in the determination of the relative phases.]

In contrast to a quasi-two-body $CP$ violation analysis,
% the time-dependent Dalitz analysis
 a TDPA
includes measurements of interfering parameters,
% corresponding to measurements of $CP$-violating asymmetries
which are measurements of $CP$-violating asymmetries
in mixed final states.
In principle, 
these measurements allow us to determine all the relative
sizes and phases of the amplitudes $A^\kappa$ and $\overline{A}{}^\kappa$,
which are related to $\phi_2$ through an isospin
 relation~\cite{Lipkin:1991st,Gronau:1991dq} by
\begin{equation}
 e^{+2i\phi_2} = 
  \frac{\overline{A}{}^+ + \overline{A}{}^- + 2 \overline{A}{}^0}
  {A^+ + A^- + 2 A^0} \;.
\end{equation}
% They in principle allow us to
Consequently, in the limit of high statistics,
 we can constrain $\phi_2$ without discrete ambiguities.

% ============================================================
% Event reconstruction
% ============================================================
To reconstruct candidate $\bz\to\pip\pim\piz$ decays,
 we combine pairs of oppositely charged tracks with $\pi^0$ candidates.
The selection criteria for charged tracks are the same as
in the previous $\bz \to \rho^\pm\pi^\mp$ analysis~\cite{Wang:2004va}.
%
%Each charged track is required to have transverse momenta greater than
% 0.1 GeV/$c$ in the laboratory frame.
%We require the tracks to be identified as pions
% [based on information from a time-of-flight system, an aerogel
%  \v{C}erenkov counter system, and the central tracker].
%Electron-like tracks are rejected.
%
% We distinguish charged kaons from pions based on
%  a kaon (pion) likelihood $\mathcal{L}_{K(\pi)}$.
% Tracks that are positively identified as electrons are rejected.
% Photons are identified as isolated ECL clusters
% that are not matched to any charged track.
Candidate $\piz$'s are reconstructed from $\gamma$ pairs
having $M_{\gamma\gamma}$ in the range 0.1178--0.1502 GeV/$c^2$,
corresponding to $\pm 3$ standard deviations ($\sigma$) in $M_{\piz}$
resolution,
and momenta greater than 0.1 GeV/$c$ in the
laboratory frame.
We require $E_\gamma > 0.05$ (0.1) GeV
in the barrel (endcap) of the electromagnetic calorimeter~\cite{Belle},
which subtends $32^\circ$--$129^\circ$
 ($17^\circ$--$32^\circ$ and $129^\circ$--$150^\circ$) with respect to
the beam axis.
% We reconstruct $\piz$ candidates from pairs of photons
%  detected in the barrel (end-cap) region with $E_\gamma > 0.05$ (0.1) GeV,
%  where $E_\gamma$ is the photon energy.
% Photon pairs with momenta greater than 0.1 GeV/$c$ in the
%  laboratory frame and
%  with an invariant mass between 0.1178 GeV/$c^2$ to 0.1502 GeV/$c^2$,
%  roughly corresponding to $\pm 3\sigma$ of the mass resolution,
%  are used as $\piz$ candidates.
%
%
%
Candidate $B$ mesons are reconstructed using two variables
 calculated in the center-of-mass frame:
 % the beam-energy constrained mass calculated using the beam energy in place
 the $B$ invariant mass calculated using the beam energy in place
 of the reconstructed energy ($\mbc$), and the energy difference between
 the $B$ candidate and the beam energy ($\dE$).
We define a signal region
 $-0.1\,\mathrm{GeV} < \dE < 0.08\,\mathrm{GeV}$
 and $5.27\,\mathrm{GeV}/c^2 < \mb$,
and a large fitting region
$|\dE| < 0.2\,\mathrm{GeV}$ and $5.2\,\mathrm{GeV}/c^2 < \mbc$.

The procedure used to measure $\Delta t$ and to determine the flavor of the
decaying $\bz$ meson, $q_\mathrm{tag}$,
and its quality, $l$, are described elsewhere~\cite{Chen:2005dr}.
%
% Continuum suppression
%
The dominant background is
$e^+e^- \rightarrow q\overline{q}\; (q=u, d, s, c)$
continuum events. 
To distinguish these jet-like events from the spherical $B$ decay signal 
events,
% we combine a set of variables that characterize the event topology
we combine modified Fox-Wolfram moments~\cite{Abe:2003yy} and the $B$
flight angle with respect to the beam direction
into a signal (background) likelihood variable $\cal L_{\rm sig(bkg)}$
and impose requirements on the likelihood ratio
$\rsigbkg \equiv \lsig/(\lsig+\lbkg)$.
These requirements depend on the quality of flavor tagging.
When more than one candidate in the same event is found
% in the fit region,
in the large fitting region,
we select the best candidate
% $based on $M_{\gamma\gamma}$ and $\rsigbkg$.
using likelihood based on $M_{\gamma\gamma}$ and $\rsigbkg$.
%
%
% Dalitz Variables
%
After the best candidate selection,
we apply a Dalitz plot cut:
candidates are required to satisfy
$0.55\,\mathrm{GeV}/c^2 < \sqrt{s_{\pm (0)}} < 1.0 \: (0.95)\,\mathrm{GeV}/c^2$
 for at least one of $s_+$, $s_-$, or $s_0$.
In the fits below, we use square Dalitz plot variables $(m', \theta')$
for convenience,
% to better elucidate structure
%  in regions where events are densely
%  concentrated.
performing a parameter transformation on $(s_+, s_-)$~\cite{Aubert:2005sk}.
% we reconstruct
% the Dalitz variables $s_+$, $s_0$ and $s_-$
% from 1) the four momenta of the $\pi^+$ and $\pi^-$,
% 2) the helicity angle of the $\rho^0$
% (i.e., the helicity angle of the $\pi^+ \pi^-$ system),
% and a mass constraint,
% ${m_{\bz}}^2 + 2{m_{\pip}}^2 + {m_{\piz}}^2 = s_+ + s_- + s_0$.
% and 3) the relation:
% ${m_{\bz}}^2 + 2{m_{\pip}}^2 + {m_{\piz}}^2 = s_+ + s_- + s_0$.
% Note that the energy of $\pi^0$ is not explicitly used here,
% which improves the resolution of the Dalitz plot variables.
% We reject candidates that are located in
% one of the following regions in the Dalitz plot:
% $\sqrt{s_0} > 0.95$ GeV/$c^2$ and $\sqrt{s_+}>1.0$ GeV$/c^2$ and $\sqrt{s_-} > 1.0$ GeV/$c^2$;
% $\sqrt{s_0} < 0.55$ GeV/$c^2$ or $\sqrt{s_+}<0.55$ GeV$/c^2$ or $\sqrt{s_-} < 0.55$ GeV/$c^2$.
% In these regions, the fraction of $\bz\to\rho\pi$ signal is small.
%
%
%
% However,
%  radial excitations
% ($\rho(1450)$ and $\rho(1700)$)
% are the dominant $\bz \rightarrow \pi^+\pi^-\pi^0$ contributions
%  in the region with $\sqrt{s} > 1.0$ GeV$/c^2$.
% Since the amplitudes
% of the radial excitations
%  are in general independent of
% the amplitude of the $\rho(770)$,
% they are considered to be background in our analysis;
%  vetoing the high mass region considerably reduces
% the systematic uncertainties due to their contribution.
%
%
%

\begin{figure}[b]
 \includegraphics[width=0.48\textwidth]{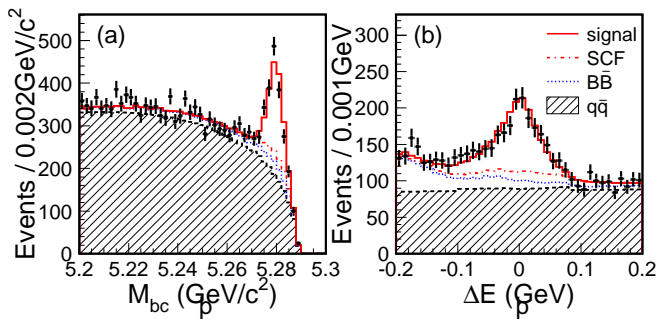}
 \caption{(a) $\mbc$ and (b) $\dE$ distributions
 within the $\dE$ and $\mbc$ signal regions.
 % The histograms are cumulative.
 Solid, dot-dashed, dotted, and dashed hatched
 histograms correspond to correctly reconstructed signal,
 SCF, $B\overline{B}$, and continuum PDF's, respectively.
 }
 \label{fig:mbc_and_de_plots}
\end{figure}

%
% Signal extraction
%
Figure~\ref{fig:mbc_and_de_plots} shows the $\mbc$ ($\dE$)
 distribution for the reconstructed
$B^0 \rightarrow \pi^+\pi^-\pi^0$ candidates
 within the $\dE$ ($\mbc$) signal region.
The signal yield is determined from an unbinned four-dimensional
extended-maximum-likelihood fit to the $\dE$-$\mbc$
and Dalitz plot distribution
 in the large fitting region,
% in the fit region defined as
%$\mbc > 5.2\,\mathrm{GeV}/c^2$ and $|\dE| < 0.2\,\mathrm{GeV}$,
where the Dalitz plot distribution is fitted only for events
inside the $\dE$-$\mbc$ signal region.
The fit function includes signal; incorrectly reconstructed signal,
 which we call self-cross-feed (SCF);
continuum; and $B\overline{B}$ background components.
The probability density function (PDF) for each component
is the same as that used for
% time-dependent analysis, 
 the TDPA described below,
% but integrated (summed) over $\Delta t$ ($q_\mathrm{tag}$).
but integrated over $\Delta t$ and summed over $q_\mathrm{tag}$.
% The $\dE$-$\mbc$ distribution of signal is modeled with binned histograms
% obtained from MC,
% where the correlation between $\dE$ and $\mbc$,
%  the dependence on $p_{\pi^0}$,
% and the difference between data and MC are taken into account.
% We also take into account incorrectly reconstructed signal events,
%  which we call self cross feed (SCF)
%  and which amounts to $\sim 20\%$ of the signal.
% In a SCF event, one of the three pions in $\fCP$ is swapped
%  with a pion in $f_\mathrm{tag}$ or the $\pi^0$ in $\fCP$ is misreconstructed.
% We give the details of the $\dE$-$\mb$ and Dalitz plot PDFs of the SCF events
%  in appendix \ref{sec:appendix_pdf_definitions}.
% For continuum,
% we use the ARGUS parameterization~\cite{Albrecht:1990am}
% for $\mbc$
% and a linear function for $\dE$.
% The $\dE$-$\mbc$ distribution of $B\overline{B}$ background
% is modeled by binned histograms based on MC.
% The Dalitz plot distributions for all components
% are modeled in the same way as the time-dependent fit
% described later,
% but integrated over the dimensions of the proper time difference,
% $\Delta t$, and the flavor of the tag side $B$, $q_\mathrm{tag}$.
%
The fit yields $971 \pm 42$
% 
% REVISED_IN_V0.7
%
% The fit to a $\lint\,\mathrm{fb}^{-1}$ data sample yields $971 \pm 42$
%
$B^0 \rightarrow \pi^+\pi^-\pi^0$ events in the signal region,
where the errors are statistical only.
%
% DE-Mbc figures
%

% ============================================================
% Lineshape determination
% ============================================================
Using the same data sample as described above
 but performing a time-integrated Dalitz plot fit
  with a wider Dalitz plot acceptance,
 $0.0 \: (0.55)\,\mathrm{GeV}/c^2 < \sqrt{s_{\pm (0)}} < 1.5\,\mathrm{GeV}/c^2$,
 we determine
 the $\rho$ lineshape
%  , i.e., the phases and amplitudes of
%  the complex coefficients $\beta$ and $\gamma$.
parameters $\beta$ and $\gamma$.
% in equation (\ref{equ:fpi_lineshape_beta_gamma}).
We use the results obtained for
% the time-dependent Dalitz plot analysis
 the TDPA
  below.
We also put upper limits on the possible deviations of
($\kakkoOverlineBeta {}_\kappa$, $\kakkoOverlineGamma {}_\kappa$)
from the nominal ($\beta$, $\gamma$),
which we use to estimate systematic errors.

To determine the 26 coefficients,
we define the following event-by-event PDF:
% \begin{equation}
\[
 P = \sum_{X = \mathrm{sig}, q\overline{q}, B\overline{B}}
  f_X
  \mathcal{P}_X(\dE, \mbc, m', \theta',\Delta t, q_\mathrm{tag}, l) \;,
\]
% \end{equation}
% \begin{widetext}
%  \begin{equation}
%   \label{equ:dembc_dalitz_simultaneous-total_pdf}
%   \begin{split}
%    P(\dE, \mbc, m', \theta', & \Delta t, q_\mathrm{tag}, l)
%     =  f_\mathrm{sig} \;
%    \frac{1}{\displaystyle n_\mathrm{true} + \sum_{i = \mathrm{NR,CR}} n_i}
%    \Bigl[
%    \mathcal{F}_\mathrm{true}^l
%    \mathcal{P}_\mathrm{true}^{K}(\dE, \mbc)
%    \epsilon (m',\theta'; l)
%    \mathcal{P}_\mathrm{true}^{D}(m',\theta', \Delta t, q_\mathrm{tag}; l)
%    \\
%    & \qquad + \!\! \sum_{i=\mathrm{CR, NR}} \!\!
%    \mathcal{F}_i^l
%    \mathcal{P}_i^{K}(\dE, \mbc; m', \theta')
%    \int \! dm'_t d\theta'_t R_i(m', \theta'; m'_t, \theta'_t)
%    % \int  
%    \mathcal{P}_i^{D}(m'_t,\theta'_t, \Delta t, q_\mathrm{tag}; l)
%    \Bigr]
%   \\
%    & + f_{q\overline{q}} \;
%    \mathcal{F}_{q\overline{q}}^l
%    \mathcal{P}_{q\overline{q}}^{K_1}(\dE; l)
%    \mathcal{P}_{q\overline{q}}^{K_2}(\mbc)
%    \mathcal{P}_{q\overline{q}}^{D}(m', \theta'; \dE, \mbc)
%    \frac{1 - q_\mathrm{tag} \mathcal{A}_{q\overline{q}}(m', \theta', l)}{2}
%    \mathcal{P}_{q\overline{q}}^{\Delta t}(\Delta t)
%   \\
%    & + f_{B\overline{B}}
%    \sum_{j \in (\mathrm{all}\;B\overline{B}\;\mathrm{modes})}
%    \mathcal{F}_j^l
%    \mathcal{P}_j^{K}(\dE, \mbc)
%    \sum_{q_\mathrm{rec}}
%    \mathcal{P}_j^{D}(m', \theta'; q_\mathrm{rec})
%    \mathcal{P}_j^{\Delta t}(\Delta t, q_\mathrm{tag}, q_\mathrm{rec}; l)
%    % & &
%    \;,
%   \end{split}
%  \end{equation}
where
$\mathcal{P}_\mathrm{sig}$, $\mathcal{P}_{q\overline{q}}$, and
$\mathcal{P}_{B\overline{B}}$ are the PDF's of signal including SCF,
continuum, and $B\overline{B}$ components,
respectively,
and
 $f_\mathrm{sig}$, $f_{q\overline{q}}$, and
 $f_{B\overline{B}}$ are the corresponding fractions that satisfy
 $f_\mathrm{sig} + f_{q\overline{q}} + f_{B\overline{B}}=1$.
Here, $\mathcal{P}_\mathrm{sig}$ and $\mathcal{P}_{B\overline{B}}$
are modeled based on Monte Carlo (MC), though a small correction is
applied
to $\mathcal{P}_\mathrm{sig}$ to take account of the difference between
data and MC, while $\mathcal{P}_{q\overline{q}}$ is modeled using data.
The signal PDF, $\mathcal{P}_\mathrm{sig}$, is the sum of
a correctly reconstructed PDF ($\mathcal{P}_\mathrm{true}$)
and an SCF PDF,
%  ($\mathcal{P}_\mathrm{SCF}$), and their normalizations
% ($n_\mathrm{true}$ and $n_\mathrm{SCF}$) as
% \begin{equation}
%  \mathcal{P}_\mathrm{sig}
%   = \frac{1}{n_\mathrm{true} + n_\mathrm{SCF}}
%   \left( \mathcal{P}_\mathrm{true} + \mathcal{P}_\mathrm{SCF} \right) \;,
% \end{equation}
where
% \begin{equation}
\[
 \begin{split}
  \mathcal{P}_\mathrm{true}
  = & \mathcal{F}^l_\mathrm{true} \mathcal{P}_\mathrm{true}(\dE, \mbc)
  \epsilon(m', \theta'; l)
 \\
  & \times
  \mathcal{P}_\mathrm{true}(m', \theta', \Delta t, q_\mathrm{tag}; l) \;.
 \end{split}
\]
% \end{equation}
Here
$\mathcal{F}^l_\mathrm{true}$, $\mathcal{P}_\mathrm{true}(\dE, \mbc)$,
$\epsilon(m', \theta'; l)$,
and $\mathcal{P}_\mathrm{true}(m', \theta', \Delta t, q_\mathrm{tag}; l)$
are
 event fractions in each category of tagging quality $l$,
 a $\dE$-$\mbc$ PDF, a Dalitz plot dependent efficiency,
 and a Dalitz-$\Delta t$ PDF 
 for the correctly reconstructed signal component,
 respectively.
The Dalitz-$\Delta t$ PDF
 corresponds to the right-hand side of 
Eq.~(\ref{equ:amplitude_dt_width})
with the following modifications:
(i) it is convolved with the
 $\Delta t$ resolution function~\cite{Tajima:2003bu};
(ii) it is multiplied by the determinant of the Jacobian
for the transformation $(s_+, s_-) \mapsto (m', \theta')$;
and
(iii) the wrong tag fractions, $w_l$, and the difference between
$\bz$ and $\bzb$ decays, $\dwl$,
are taken into account.
A more detailed description of the PDF can be found
 elsewhere~\cite{Abe:2006yg}.

\begin{figure}[b]
 \includegraphics[width=0.48\textwidth]{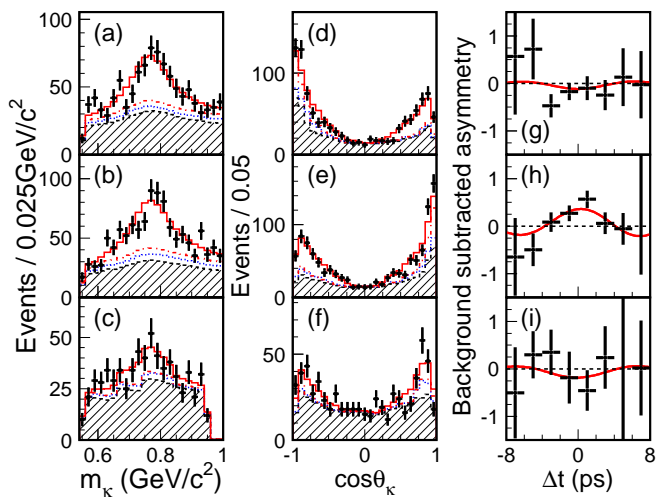}
 \caption{
 Mass (a)--(c) and helicity (d)--(f) distributions,
 and background subtracted $\Delta t$ asymmetry plots
 in the good tagging quality region $l \geq 3$~\cite{Chen:2005dr} (g)-(i),
 corresponding to the
 $\rho^+\pi^-$ [(a),(d),(g)],
 $\rho^-\pi^+$ [(b),(e),(h)], and
 $\rho^0\pi^0$ [(c),(f),(i)] enhanced regions.
 The notations for histograms (a)--(f)
 are the same as Fig.~\ref{fig:mbc_and_de_plots}.%
\label{fig:mass_helicity_dt_plot}%
 }
\end{figure}

An unbinned-maximum-likelihood fit to the 2824 events in the signal
region yields
the results listed in Table~\ref{tbl:dt_all_data}.
\begin{table*}[t]
\caption{Results of the time-dependent Dalitz fit (left three columns),
 and the associated quasi-two-body $CP$ violation parameters
 (rightmost column),
 whose definitions can be found elsewhere~\cite{Wang:2004va}.
 The first and second errors are
 statistical and systematic, respectively.
 The correlation coefficient between
 $\mathcal{A}_{\rho\pi}^{+-}$ and $\mathcal{A}_{\rho\pi}^{-+}$
 ($\mathcal{A}_{\rho^0\pi^0}$ and $\mathcal{S}_{\rho^0\pi^0}$)
 is $+0.47$ ($-0.08$).\label{tbl:dt_all_data}%
 }
\begin{tabular}
{@{\hspace{2mm}}l@{\hspace{2mm}}c@{\hspace{2mm}}|@{\hspace{2mm}}l@{\hspace{2mm}}c@{\hspace{2mm}}|@{\hspace{2mm}}l@{\hspace{2mm}}c@{\hspace{2mm}}c@{\hspace{6mm}}c@{\hspace{2mm}}c@{\hspace{2mm}}}
% \hline \hline
\cline{1-6} \cline{8-9} &&&&&&&& \vspace{-2.5ex} \\ \cline{1-6} \cline{8-9}
$U^+_+$                   &  $+1$ (fixed)                & $U^-_+$                   & $+0.23 \pm 0.15 \pm 0.07$ & $I_+$                     & $-0.01 \pm 0.11 \pm 0.04$ && $\mathcal{A}_{\rho\pi}^{CP}$ & $-0.12 \pm 0.05 \pm 0.04$ \\
$U^+_-$                   & $+1.27 \pm 0.13 \pm 0.09$ & $U^-_-$                   & $-0.62 \pm 0.16 \pm 0.08$ & $I_-$                     & $+0.09 \pm 0.10 \pm 0.04$ && $\mathcal{C}$                & $-0.13 \pm 0.09 \pm 0.05$ \\
$U^+_0$                   & $+0.29 \pm 0.05 \pm 0.04$ & $U^-_0$                   & $+0.15 \pm 0.11 \pm 0.08$ & $I_0$                     & $+0.02 \pm 0.09 \pm 0.05$ && $\Delta \mathcal{C}$         & $+0.36 \pm 0.10 \pm 0.05$ \\
$U^{+, \mathrm{Re}}_{+-}$ & $+0.49 \pm 0.86 \pm 0.52$ & $U^{-, \mathrm{Re}}_{+-}$ & $-1.18 \pm 1.61 \pm 0.72$ & $I^{\mathrm{Re}}_{+-}$    & $+1.21 \pm 2.59 \pm 0.98$ && $\mathcal{S}$                & $+0.06 \pm 0.13 \pm 0.05$ \\
$U^{+, \mathrm{Re}}_{+0}$ & $+0.29 \pm 0.50 \pm 0.35$ & $U^{-, \mathrm{Re}}_{+0}$ & $-2.37 \pm 1.36 \pm 0.60$ & $I^{\mathrm{Re}}_{+0}$    & $+1.15 \pm 2.26 \pm 0.92$ && $\Delta \mathcal{S}$         & $-0.08 \pm 0.13 \pm 0.05$ \\
\cline{8-9}
$U^{+, \mathrm{Re}}_{-0}$ & $+0.25 \pm 0.60 \pm 0.33$ & $U^{-, \mathrm{Re}}_{-0}$ & $-0.53 \pm 1.44 \pm 0.65$ & $I^{\mathrm{Re}}_{-0}$    & $-0.92 \pm 1.34 \pm 0.80$ && $\mathcal{A}_{\rho\pi}^{+-}$ & $+0.21 \pm 0.08 \pm 0.04$ \\
$U^{+, \mathrm{Im}}_{+-}$ & $+1.18 \pm 0.86 \pm 0.34$ & $U^{-, \mathrm{Im}}_{+-}$ & $-2.32 \pm 1.74 \pm 0.91$ & $I^{\mathrm{Im}}_{+-}$    & $-1.93 \pm 2.39 \pm 0.89$ && $\mathcal{A}_{\rho\pi}^{-+}$ & $+0.08 \pm 0.16 \pm 0.11$ \\
\cline{8-9}
$U^{+, \mathrm{Im}}_{+0}$ & $-0.57 \pm 0.35 \pm 0.51$ & $U^{-, \mathrm{Im}}_{+0}$ & $-0.41 \pm 1.00 \pm 0.47$ & $I^{\mathrm{Im}}_{+0}$    & $-0.40 \pm 1.86 \pm 0.85$ && $\mathcal{A}_{\rho^0\pi^0}$  & $-0.49 \pm 0.36 \pm 0.28$ \\
$U^{+, \mathrm{Im}}_{-0}$ & $-1.34 \pm 0.60 \pm 0.47$ & $U^{-, \mathrm{Im}}_{-0}$ & $-0.02 \pm 1.31 \pm 0.83$ & $I^{\mathrm{Im}}_{-0}$    & $-2.03 \pm 1.62 \pm 0.81$ && $\mathcal{S}_{\rho^0\pi^0}$  & $+0.17 \pm 0.57 \pm 0.35$ \\
\cline{1-6} \cline{8-9}
\end{tabular}
\end{table*}
%
% With a toy Monte Carlo (MC) study,
With a toy MC study,
we find that the errors estimated by the likelihood function
do not give correct 68.3\% confidence level (C.L.)
 coverage for the interfering parameters.
In the table, we multiply the error estimates from the likelihood
function
% by a factor 1.19,
 by a factor of 1.17,
which is calculated from the MC study,
to obtain errors with correct coverage.
%
% We observe non-zero $U^+_0$ with significance of
% $4.8\,\sigma$, corresponding to clear evidence for the
% $\bz \to \rho^0\pi^0$ decay channel;
% [this confirms our previous measurement~\cite{Dragic:2006yv}].
%
We find that $U^+_0$ is $4.8\,\sigma$ above zero,
corresponding to clear evidence for the
presence of the decay $\bz \to \rho^0\pi^0$
in agreement with our previous measurement~\cite{Dragic:2006yv}.
%
%  (, and give the products as the errors for these parameters.
%  [Maybe to be removed. But I thought it is better to make it clear that
%  the errors listed in the table is not the "raw" errors but the corrected
%  errors. Is it evident from the context?])
%
%
% The correlation matrix of the 26 parameters
% after combining statistical and systematic errors
% is shown in the appendix \ref{sec:appendix_correlation}.
% Figure \ref{fig:dalitz_plot}
% shows the projections of the square Dalitz plot in data
% with the fit result superimposed.
%
Figure~\ref{fig:mass_helicity_dt_plot}
shows the mass and helicity distributions,
and the background-subtracted $\Delta t$ asymmetry plot
for each $\rho \pi$ enhanced region.
%
% Figure~\ref{fig:mass_helicity_dt_plot}
% shows the mass and helicity distributions
% for each $\rho \pi$ enhanced region;
% and the background-subtracted $\Delta t$ asymmetry plot
% for each region is also shown, under tighter cuts on the flavor-tagging
% quality ($l=3,4,5,6$; for display purposes only).
%
% Figure \ref{fig:dt_plot_result}
% shows the $\Delta t$ distributions and background 
% subtracted asymmetries.
We define the asymmetry in each $\Delta t$ bin by
 $(N_+ - N_-) / (N_+ + N_-)$,
where $N_{+(-)}$ corresponds to the
background-subtracted number of events with $q_\mathrm{tag}=+1\:(-1)$.
The $\rho^- \pi^+$ enhanced region shows a significant
% asymmetry proportional to $\cos (\dmd \Dt)$,
 cosine-like asymmetry
arising from a non-zero value of $U^-_-$.
Note that this is not a $CP$-violating effect,
since $\rho^-\pi^+$ is not a $CP$-eigenstate.
No sine-like asymmetry is observed in any of the regions (g)--(i).
% No $\sin$-like asymmetry is observed in any of the three plots;
% the values of $\mathcal{S}$, $\Delta \mathcal{S}$,
% and $\mathcal{S}_{\rho^0\pi^0}$ in Table~\ref{tbl:dt_all_data} are
% consistent with zero.

The non-interfering parameters can be interpreted as
 the quasi-two-body parameters of the process $\bz \to \rho^\pm\pi^\mp$,
 whose definitions can be found elsewhere~\cite{Wang:2004va},
 and the $CP$ violation parameters of the process $\bz \to \rho^0\pi^0$:
 $\mathcal{A}_{\rho^0\pi^0} = -U^-_0/U^+_0$
 and $\mathcal{S}_{\rho^0\pi^0} = 2I_0/U^+_0$.
 These are also listed in the Table~\ref{tbl:dt_all_data}.
There are several sources of systematic uncertainty.
To determine their magnitudes,
 we vary each possible contribution to the systematic error
 by its uncertainty in the data fit or in the MC,
 and take the resultant deviations
 in the fitted parameters as errors.
We add each contribution in quadrature to obtain the total systematic
 uncertainty.
The largest contribution for the interfering parameters
comes from radial excitations.
We take account of
% the associated uncertainties as follows:
possible deviations of
 ($\kakkoOverlineBeta {}_\kappa$, $\kakkoOverlineGamma {}_\kappa$)
 from the ($\beta$, $\gamma$) values,
% used for the nominal fit,
% and uncertainties of the mass and width of each resonance, and
% 3) uncertainties in $\beta$ and $\gamma$.
and uncertainties of $\beta$, $\gamma$, and the mass and width of each resonance.
Large contributions to the systematic errors
for the non-interfering parameters
come from potential backgrounds such as
%  the $B^0 \rightarrow \pi^+\pi^-\pi^0$
% decay processes that are not $B^0 \rightarrow (\rho\pi)^0$.
% We take account of the contributions from
$B^0 \rightarrow f_0(980) \pi^0, f_0(600) \pi^0, \omega \pi^0$,
and non-resonant $\pipipi$, which we neglect in our nominal fit.
We perform fits to toy MC including these backgrounds with the branching
fractions 
at their 68.3\% C.L. upper limits,
which we obtain from our data or world
averages~\cite{Eidelman:2004wy,unknown:2006bi};
the largest variations are taken as systematic errors.
%
% [Upper limits of one standard deviation ($\sigma$)
% are set on their branching fractions using our data or quoting world
%  averages~\cite{Eidelman:2004wy,unknown:2006bi};
% assuming the 1$\sigma$ upper limit values
%  for the branching fractions,
% we perform toy MC studies to estimate the largest possible
% impacts on our measurement.]
%
% Upper limits on their contributions
%  are determined from data,
% except for $B^0 \rightarrow \omega \pi^0$,
% for which we use the world average of
%  $\mathcal{B}(B^0 \rightarrow \omega \pi^0)$~\cite{unknown:2006bi}
% and $\mathcal{B}(\omega \rightarrow \pi^+\pi^-)$~\cite{Eidelman:2004wy}.
% We use recent measurements~\cite{Ablikim:2004qn,Muramatsu:2002jp,Aitala:2000xu}
% for the mass and width parameters of the $f_0(600)$ resonance.
% We find no significant signals for any of the above decay modes.
% %
% Using the $1\sigma$ upper limits as input,
% we generate toy MC
% for each mode
% with the interference between
% the $B^0 \rightarrow (\rho \pi)^0$
% and the other
% $B^0 \rightarrow \pi^+ \pi^- \pi^0$ mode taken into account.
% We obtain the systematic error by
% fitting the toy MC assuming
% $B^0 \rightarrow (\rho \pi)^0$ only in the PDF.
%
%%%%%%%%%
% Recover in 0.8
Comparable contributions also come from vertex reconstruction,
background PDF's, and tag-side interference~\cite{Long:2003wq};
more detail can be found elsewhere~\cite{Abe:2006yg}.
%%%%%%%%%
% OMIT_IN_V0.7
%  Systematic errors related to the vertexing method
%  and tag-side interference~\cite{Long:2003wq}
%  also give large contributions;
%  they are estimated in the same manner as in Ref.~\cite{Chen:2005dr}.
%%%%%%%%%

% We observed fit bias due to small statistics for some of the fitted parameters.
% Since this bias is much smaller than the statistical error,
% we take it into account in the systematic errors.
% A large fit bias is observed with our nominal fitted parameters,
% which is mainly due to small statistics.
% The dominant source of the bias
% is found to be an intrinsic bias due to small statistics.
% We estimate the size of the fit bias by Toy MC study
% and quote the bias as the systematic errors.
% We also confirm that the bias
% is consistent between Toy MC and full detector MC simulation.

% Finally, we investigate the effects of tag-side interference (TSI),
% which is the interference between
% CKM-favored and CKM-suppressed $B\rightarrow D$ transitions
% in the $f_\mathrm{tag}$ final state~\cite{Long:2003wq}.
% A small correction to the PDF for the signal distribution arises
% from the interference.
% We estimate the size of the correction using the
%  $\bz \rightarrow D^{*-}\ell^+ \nu$ sample.
% We then generate MC pseudo-experiments and make an 
% ensemble test to obtain the systematic biases.

\begin{figure}[b]
 \begin{center}
  \includegraphics[width=0.42\textwidth]{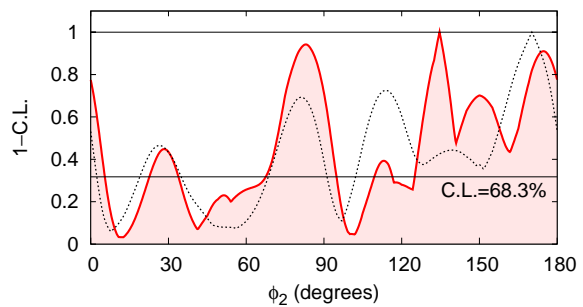}
  \caption{
  $1-\mathrm{C.L.}$ vs. $\phi_2$.
  Dotted and solid curves
  correspond to the result from 
  % the time-dependent Dalitz plot analysis
  the TDPA
  only and that from 
  % the Dalitz
  the TDPA
  and an isospin (pentagon) combined analysis,
  respectively.
  \label{fig:phi2_1_cl_plot}
  }
 \end{center}
\end{figure}

% ============================================================
% phi2
% ============================================================
We constrain $\phi_2$ from the 26 parameters
 measured in our analysis
 following the formalism of Ref.~\cite{Snyder:1993mx}
 and the statistical treatment
 using toy MC described in Ref.~\cite{Charles:2004jd}.
The resulting
 $\mathrm{1-C.L.}$ function is shown in
 Fig.~\ref{fig:phi2_1_cl_plot} as a dotted curve.
To incorporate all available knowledge,
we combine our measurement with results on the
 branching fractions for $B^0 \rightarrow \rho^\pm \pi^\mp$
 and $B^+ \to \rho^+ \pi^0$, $\rho^0 \pi^+$,
 and flavor asymmetries
%  for $B^+ \rightarrow \rho^+ \pi^0$
%  and $B^+ \rightarrow \rho^0 \pi^+$~\cite{unknown:2006bi}.
 of the latter two~\cite{unknown:2006bi}.
Assuming isospin (pentagon) relations~\cite{Lipkin:1991st,Gronau:1991dq}
 and following the same procedure as above,
 we perform a full Dalitz and pentagon combined analysis,
the result of which is shown in Fig.~\ref{fig:phi2_1_cl_plot}
as the solid curve.
% and isospin relations~\cite{Lipkin:1991st,Gronau:1991dq}.
% With five $B \rightarrow \rho\pi$ decay modes,
% we have 12 free parameters including $\phi_2$:
% \begin{equation}
%  \begin{array}{l}
%   12 =
%    \mathrm{(10\;complex\;amplitudes=20 \, d.o.f.) + \phi_2}
%    \\
%   \hspace{4mm}
%    \mathrm{
%    - (1\;global\;phase)
%    - (4\;isospin\;relations=8 \, d.o.f.)} \;.
%  \end{array}
% \end{equation}
% Parameterizing the 10 complex amplitudes with 12 free parameters,
% we form a $\chi^2$ function using as constraints the 26 measurements
% of our analysis
% and the
% world average branching fractions and asymmetries:
% $\mathcal{B}(B^0 \rightarrow \rho^\pm \pi^\mp)$,
% $\mathcal{B}(B^+ \rightarrow \rho^+ \pi^0)$,
% $\mathcal{A}(B^+ \rightarrow \rho^+ \pi^0)$,
% $\mathcal{B}(B^+ \rightarrow \rho^0 \pi^+)$,
% and
% $\mathcal{A}(B^+ \rightarrow \rho^0 \pi^+)$~\cite{unknown:2006bi},
% which are not correlated with our 26 observables.
% With the $\chi^2$,
% we perform a full combined
% Dalitz and isospin(pentagon) analysis
% following the toy MC procedure
% described in the Ref.~\cite{Charles:2004jd},
%
We obtain $68^\circ < \phi_2 < 95^\circ$ as the 68.3\% confidence
interval for the solution consistent with the SM expectation.
A large SM-disfavored region
 ($0^\circ<\phi_2<5^\circ$, $23^\circ <\phi_2<34^\circ$,
 and $109^\circ <\phi_2<180^\circ$) also
 remains.
% AHO
% We obtain $\phi_2 = (83^{+12}_{-23})^\circ$ as the central value
% and $1\sigma$ errors (corresponding to 68.3\% C.L.).
% A large CKM-disfavored region
%  ($\phi_2<8^\circ$ and $129^\circ<\phi_2$) 
% also remains.
In principle, with more data we may be able to remove the additional
$\phi_2$ solutions.

% ============================================================
% Summary
% ============================================================
In summary,
using $\lint \, \mathrm{fb}^{-1}$ of data
we have performed a full Dalitz plot analysis
of the $B^0 \rightarrow \pi^+\pi^-\pi^0$ decay mode,
where the observables include
the first measurement of $\mathcal{S}_{\rho^0\pi^0}$.
A full
 time-dependent Dalitz plot analysis
 with the pentagon isospin
relation is performed for the first time
and a constraint on the angle $\phi_2$ is obtained.
% combining our analysis with information
% from charged $B$ decay modes,
% we obtain $ = (83^{+12}_{-23})^\circ$ as the central value
% with $1\sigma$ errors.

% ============================================================
% Acknowledge
% ============================================================
%-------- Short version, if necessary, for PRL -----------
We thank the KEKB group for excellent operation of the
accelerator, the KEK cryogenics group for efficient solenoid
operations, and the KEK computer group and
the NII for valuable computing and Super-SINET network
support.  We acknowledge support from MEXT and JSPS (Japan);
ARC and DEST (Australia); NSFC and KIP of CAS (China); 
DST (India); MOEHRD, KOSEF and KRF (Korea); 
KBN (Poland); MIST (Russia); ARRS (Slovenia); SNSF (Switzerland); 
NSC and MOE (Taiwan); and DOE (USA).

\bibliographystyle{apsrev.bst}
\bibliography{cites}

\begin{thebibliography}{22}
\bibitem[{Cha()}]{ChargeConjugate}
\bibinfo{note}{Throughout this Letter, the inclusion of the charge conjugate
  decay mode is implied unless otherwise stated.}
\bibitem[{Phi()}]{Phi2Alpha}
\bibinfo{note}{Another naming convention, $\alpha$ ($= \phi_2$), is also used
  in the literature.}
\end{thebibliography}

\end{document}